# Spectrophotometric Indices and Metal Content of Galactic Globular Clusters

S. Covino[1], S. Galletti[1], and L.E. Pasinetti[1]

Dip. di Fisica, Univ. di Milano, via Celoria 16, I-20133, Milano, Italy

**Abstract.** Spectrophotometric indices for 18 Galactic globular clusters, obtained from CCD observations and careful reductions, were used to determine reliable calibrations on metallicity $[Fe/H]$. The indices were measured in the bandpasses adopted by Burnstein et al. (1984). Adding other observations of Burnstein et al. (1984) we obtained our results from an homogeneous sample of indices for 26 globular clusters.

Relations with indices defined by other Authors and with metallicity photometric indices or parameters were also computed. In each case the relations are quite satisfactory.

Observational data were compared with synthetic indices derived from Buzzoni's (1989) models and detailed discussions were performed for $Mg_2$, $Fe_{52}$, and $H_\beta$. The observational points seem to be systematically shifted with respect to the fiducial lines traced by the models. The scenario confirms that a certain degree of oxygen enhancement would be necessary to obtain a better agreement between observed data and theoretical predictions. This enhancement, however, removes some of the disagreement, but not all of it. The dependence of the observed $Fe_{52}$ and $H_\beta$ indices on the metal content for different HB morphologies was considered.

Finally, some results were also discussed from a statistical point of view. A principal component analysis was applied to the index sample to study the number of independent parameters necessary to reproduce the observations. The whole index set is completely consistent with a one-parameter family.

**Key words:** globular clusters: general – Galaxy: abundances – Galaxy: halo – Galaxy: stellar content

## 1. Introduction

Globular Cluster (GC) systems offer an unique opportunity to test simultaneously stellar evolutionary theories and models of galactic formation and evolution. Stellar populations studies are also strictly related to the understanding of the features of the most simple and pure stellar populations available to observations.

One of the most important parameters to know about GCs is their metal content even if it is not always well defined what is meant by the term "metallicity". Ideally, detailed comparison between spectra of individual stars in a GC with models of stellar atmospheres can provide a reliable estimate of the chemical abundances. This approach is, however, difficult to perform and extremely time consuming. The measure of some features in a c-m diagram gives an alternative approach and allows one to determine the metal content of a GC. In any case, it is necessary an accurate photometry of a large number of stars in each cluster.

Many authors (see for instance Burnstein et al. 1984, Armandroff & Zinn 1988, Brodie & Huchra 1990) identified a large set of features in integrated spectra of late-type stellar systems sensitive to changes in the mean chemical abundances and defined indices to measure the strengths of these features. Some of them can be combined in order to obtain valuable estimate of the metal content.

The aim of this study is to enlarge the sample of observed spectrophotometric indices defined in an homogeneous and reproducible (i.e. non instrumental) system for Galactic Globular Clusters (GGCs), to determine reliable calibrations of indices versus $[Fe/H]$ and other metallicity parameters, and to explore the theoretical predictions with respect our and previous observations. We have also performed a statistical analysis in order to assess the sensitivity of indices as metallicity indicators and to study the number of independent parameters actually measured by spectrophotometric indices for GGCs. A statistical study on GC systems was made by Covino & Pasinetti (1993) using parameters found in the literature.

| Date | Grating (l/mm) | CCD | Wavelength Range (Å) | Resolution (Å) |
|---|---|---|---|---|
| April 9-13, 1989 | 1200 | RCA SID 006 EX | 4580 − 5560 | 2.1 |
| July 5-6, 1993 | 220 | FORD 2048L | 3650 − 11000 | 8.1 |
| July 6-8, 1993 | 600 | FORD 2048L | 3700 − 7500 | 3.4 |

**Table 1.** Observation data.

In the following of this paper, we give observation data and related reductions (sect. 2), the values of metallicity indices (sect. 3), the calibrations of $[Fe/H]$ obtained for each index (sect. 4), a comparison between observations and synthetic indices determined by Buzzoni's models of Integrated Spectral Energy Distribution (ISED) (Buzzoni 1989, 1993, Buzzoni et al. 1992, 1994) (sect. 5). Finally, we performed an analysis of the statistical dimensionality for a sample of indices of GGCs (sect. 6).

[b]

## 2. Observations and Reductions

Spectra of 18 GGCs were taken at the 1.52 m spectrographic telescope of the ESO La Silla observatory, equipped with a B&C spectrograph and both a RCA CCD (1024 × 640 pixels) and a Ford CCD (2048 × 2048 pixels). Observation data are reported in Table 1 and the list of the observed GCs in Table 3. The clusters were selected with the criterion to cover the full range of metallicity spanned by GGCs and a wide range of physical parameters as central concentrations, $M/L$ ratio, etc.

Each spectrum was reduced in standard way by the ESO-MIDAS package slightly modified in some procedures. Data obtained on July 5-8, 1993 were also flux calibrated by means of spectrophotometric standard stars. In order to obtain spectra representative of each cluster, but those with the smallest core radii, we averaged light from the central region moving the telescope perpendicularly to the slit during the exposure (Armandroff & Zinn 1988) or, alternatively, averaging spectra of different zone of each cluster. The region sampled was always of the order of the core radius, or larger.

## 3. Measure of the Indices

Many authors have defined line indices as tools to measure metallicity, temperature and surface gravity from medium-low resolution spectra of stars. The same indices were used as metallicity indicators in integrated spectra of stellar systems. Each line index is a measurement of the flux contained in a wavelength region centered on the feature relative to that contained in red and blue pseudo-continuum regions close to the feature (Faber et al. 1984). We have determined the indices following the prescriptions and formulas given by Faber et al. (1977). If $F(\lambda_1, \lambda_2)$ is the average instrumental flux per unit wavelength in the $(\lambda_1, \lambda_2)$ interval, we compute the index $I$ as:

$$I = -2.5 \log\left(\frac{F_c}{F_b + k \cdot (F_r - F_b)}\right), \quad (1)$$

where $c, b$, and $r$ refer to the central, blue, and red bandpasses. $k$ is a constant defined by an interpolation between the central wavelengths in the blue and red bandpasses with respect the central wavelength of the bandpass centered on the feature:

$$k = \frac{(\lambda_{F_{c_2}} - \lambda_{F_{b_2}}) - (\lambda_{F_{c_1}} - \lambda_{F_{b_1}})}{(\lambda_{F_{r_2}} - \lambda_{F_{b_2}}) - (\lambda_{F_{r_1}} - \lambda_{F_{b_1}})}. \quad (2)$$

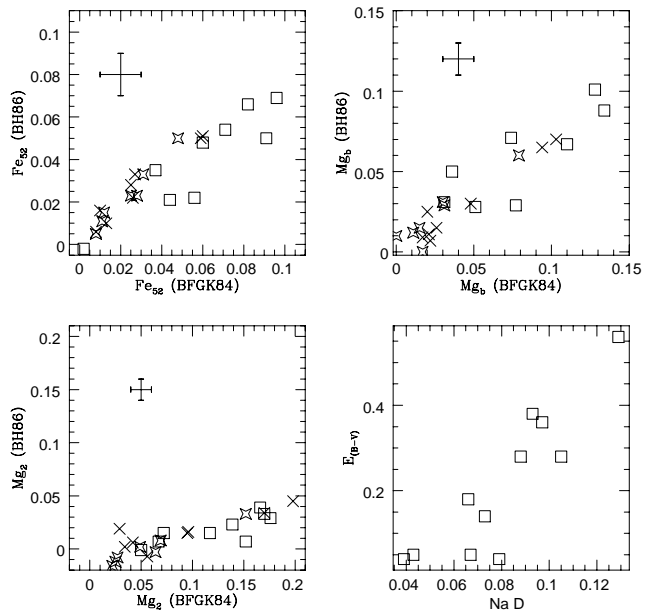

**Fig. 1.** Comparison between indices measuring the same feature but defined on different bandpasses. Squares refer to observations with a resolution of 2.1 Å, crosses of 3.4 and stars of 8.1 Å. Bottom-right panel: $Na\,D$ index vs. $E_{(B-V)}$. This feature is also sensitive to interstellar sodium line.

In this way, the denominator in the logarithm represents the pseudo-continuum at the wavelength of the feature.

Instrumental indices were determined using bandpasses reported by Tobin & Nordsieck (1981), Burstein et al. (1984) (BFGK84), Brodie & Hanes (1986) (BH86), Armandroff & Zinn (1988), Held & Mould (1994), and Silchenko (1994). The widths of central and continuum regions were chosen by these Authors to minimize the effect of typical galaxy velocity dispersions on the measured absorption-line strengths.

| Identification | Central Bandpass (Å) | Continuum Bandpasses (Å) | |
|---|---|---|---|
| $G\,4300\ (CH)$ | 4283.25-4317.00 | 4268.25-4283.25 | 4320.75-4335.75 |
| $Mg_b\,5177\ (Mg_b)$ | 5162.00-5193.25 | 5144.50-5162.00 | 5193.25-5207.00 |
| $Fe\,5270\ (Fe\,I, Ca\,I)$ | 5248.00-5286.75 | 5235.50-5249.25 | 5288.00-5319.25 |
| $Fe\,5335\ (Fe\,I, Cr\,I, Ca\,I, Ti\,II)$ | 5314.75-5353.50 | 5307.25-5317.25 | 5356.00-5364.75 |
| $Na\,5895\ (Na\,D)$ | 5879.25-5910.50 | 5863.00-5876.75 | 5924.50-5949.25 |
| $H_\beta\,4861\ (H_\beta)$ | 4849.50-4877.00 | 4829.50-4848.25 | 4878.25-4892.00 |
| $CN\ (CN)$ | 4144.00-4177.75 | 4082.00-4118.25 | 4246.00-4284.75 |
| $Mg_1\ (MgH)$ | 5071.00-5134.75 | 4897.00-4958.25 | 5303.00-5366.75 |
| $Mg_2\ (MgH+Mg_b)$ | 5156.00-5197.25 | 4897.00-4958.25 | 5303.00-5366.75 |

**Table 2.** Index definitions as reported by BFGK84.

As pointed out by Rose (1984), spectral indices are relative measures, calibrated on spectra of standard stars with known atmospheric parameters. Following the works of Faber et al. (1977, 1984) we have converted the instrumental indices into a homogeneous system by a comparison with a suitable number of standard stars. A large database, existing only for indices defined according to Faber et al. (1977), is reported in BFGK84 (Buzzoni et al. 1992, 1994). Therefore we have considered in this study only indices computed in the BFGK84 bandpasses (Table 2). It is easily possible to convert index values from magnitudes to pseudo-equivalent widths and vice versa by the formulas:

$$EW(I) = (\lambda_2 - \lambda_1)(1 - 10^{-\frac{I}{2.5}}) \quad (3)$$

$$I(EW) = -2.5\log(1 - \frac{EW}{\lambda_2 - \lambda_1}). \quad (4)$$

All the instrumental indices are reported by Covino (1995) and are available upon request by the Authors together with the final reduced spectra.

In the index determination, particular care was devoted to quantify various possible errors. Among error sources we considered effects of wavelength calibration, radial velocity correction, sky subtraction, flux calibration, and statistical Poissonian noise. As it was expected, the Poissonian fluctuation on photon number, derived according to BH90, is the largest error source depending on the $S/N$ of the final spectrum, as shown also by adequate simulations. Different observations of the same object were also compared. Table 3 reports indices of the observed GCs computed according to Faber et al. (1977) system. The values are given in magnitude units. The error bars reported in the table and in figures are deduced taking into account the statistical noise and the effect of the calibration procedure.

Figure 1 shows the correlations between values of the same instrumental index, defined on different bandpasses by BFGK84 and BH86. As we can see, the relations are satisfactory; the range of these values, however, may be different. The indices obtained from different resolution spectra are quite comparable except the case of the index $Fe_{52}$ for which slightly different correlations can be obtained from $2.1\,\text{Å}$ and $3.4, 8.1\,\text{Å}$ resolution spectra. $Na\,D$ index can be contaminated by interstellar lines and therefore gives less useful information. A panel of Fig. 1 shows its correlation with $E_{(B-V)}$ (reddening from Peterson 1993).

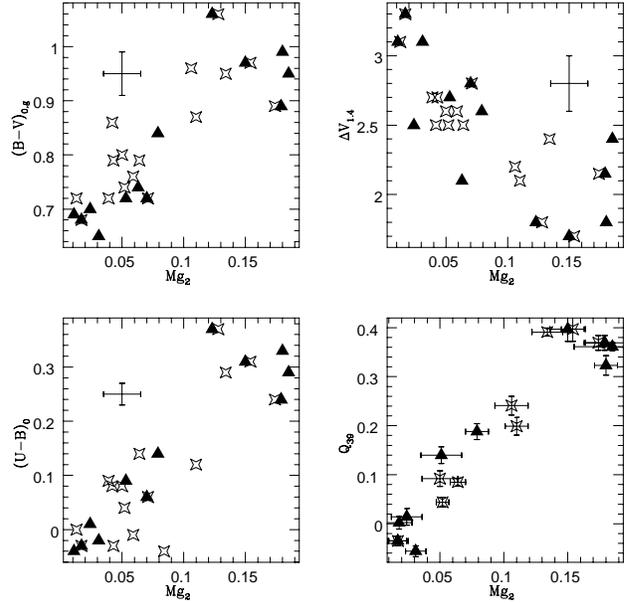

**Fig. 2.** $Mg_2$ indices from this work (triangles) and BFGK84 (crosses) vs. metallicity parameters: $(U-B)_0$ (Reed et al. 1988), $Q_{39}$ (Zinn & West 1984), $(B-V)_{0,g}$ (Webbink 1985), and $\Delta V_{1.4}$ (Peterson 1986).

| Cluster | $CH$ | $Mg_b$ | $Fe_{52}$ | $Fe_{53}$ | $NaD$ | $H_\beta$ | $CN$ | $Mg_1$ | $Mg_2$ |
|---|---|---|---|---|---|---|---|---|---|
| NGC 104 | 0.150 | 0.106 | 0.063 | 0.054 | 0.089 | 0.066 | 0.013 | 0.069 | 0.180 |
| = 47 Tuc | ±0.008 | ±0.007 | ±0.008 | ±0.007 | ±0.008 | ±0.007 | ±0.008 | ±0.008 | ±0.009 |
| NGC 362 | 0.109 | 0.002 | 0.038 | 0.027 | 0.039 | 0.066 | −0.042 | 0.028 | 0.079 |
|  | ±0.012 | ±0.008 | ±0.010 | ±0.008 | ±0.009 | ±0.009 | ±0.011 | ±0.009 | ±0.009 |
| NGC 5024 |  | 0.027 |  | 0.013 |  | 0.111 |  | 0.001 | 0.053 |
| = M53 |  | ±0.010 |  | ±0.010 |  | ±0.011 |  | ±0.012 | ±0.013 |
| NGC 5824 | 0.049 | 0.003 | 0.039 | 0.029 | 0.073 | 0.101 | −0.091 | −0.006 | 0.024 |
|  | ±0.017 | ±0.008 | ±0.011 | ±0.010 | ±0.011 | ±0.012 | ±0.014 | ±0.011 | ±0.012 |
| NGC 5946 | 0.096 | 0.019 | 0.044 | 0.040 | 0.129 | 0.082 | −0.097 | 0.022 | 0.051 |
|  | ±0.030 | ±0.009 | ±0.013 | ±0.015 | ±0.014 | ±0.013 | ±0.024 | ±0.013 | ±0.016 |
| NGC 6171 |  | 0.058 | 0.082 | 0.038 |  | 0.057 |  |  |  |
|  |  | ±0.008 | ±0.010 | ±0.009 |  | ±0.012 |  |  |  |
| NGC 6218 |  | 0.060 | 0.032 | 0.024 |  | 0.049 |  | 0.020 | 0.070 |
|  |  | ±0.010 | ±0.012 | ±0.011 |  | ±0.012 |  | ±0.009 | ±0.010 |
| NGC 6284 | 0.059 | 0.038 | 0.043 | 0.031 | 0.088 | 0.115 | −0.082 | 0.027 | 0.077 |
|  | ±0.020 | ±0.011 | ±0.014 | ±0.012 | ±0.013 | ±0.015 | ±0.017 | ±0.010 | ±0.011 |
| NGC 6293 | 0.044 | −0.004 | 0.028 | 0.016 | 0.097 | 0.105 | −0.106 | −0.005 | 0.018 |
|  | ±0.016 | ±0.009 | ±0.013 | ±0.010 | ±0.012 | ±0.015 | ±0.016 | ±0.009 | ±0.010 |
| NGC 6356 |  | 0.101 | 0.068 | 0.032 |  | 0.067 |  | 0.062 | 0.179 |
|  |  | ±0.010 | ±0.015 | ±0.011 |  | ±0.010 |  | ±0.013 | ±0.015 |
| NGC 6397 | 0.006 | 0.005 | 0.029 | 0.022 | 0.066 | 0.132 | −0.142 | −0.008 | 0.011 |
|  | ±0.023 | ±0.011 | ±0.011 | ±0.008 | ±0.009 | ±0.010 | ±0.027 | ±0.010 | ±0.011 |
| NGC 6624 | 0.165 | 0.090 | 0.061 | 0.052 | 0.105 | 0.078 | 0.020 | 0.050 | 0.150 |
|  | ±0.021 | ±0.012 | ±0.014 | ±0.012 | ±0.013 | ±0.015 | ±0.018 | ±0.012 | ±0.014 |
| NGC 6626 | 0.091 | 0.039 | 0.044 | 0.035 | 0.093 | 0.101 | −0.052 | −0.002 | 0.063 |
|  | ±0.017 | ±0.008 | ±0.014 | ±0.011 | ±0.012 | ±0.013 | ±0.016 | ±0.013 | ±0.014 |
| NGC 6637 |  | 0.097 | 0.047 | 0.044 |  | 0.055 |  | 0.050 |  |
| = M69 |  | ±0.008 | ±0.008 | ±0.006 |  | ±0.016 |  | ±0.016 |  |
| NGC 6712 |  |  | 0.043 | 0.023 |  | 0.061 |  |  |  |
|  |  | ±0.012 | ±0.020 | ±0.018 |  | ±0.018 |  |  |  |
| NGC 6838 |  |  |  |  |  | 0.069 |  | 0.033 | 0.123 |
| = M71 |  |  |  |  |  | ±0.014 |  | ±0.015 | ±0.016 |
| NGC 7078 | 0.022 | 0.011 | 0.028 | 0.015 | 0.067 | 0.105 | −0.102 | −0.005 | 0.017 |
| = M15 | ±0.007 | ±0.005 | ±0.008 | ±0.006 | ±0.007 | ±0.006 | ±0.006 | ±0.008 | ±0.008 |
| NGC 7099 | 0.027 | 0.013 | 0.025 | 0.022 | 0.043 | 0.120 | −0.115 | −0.002 | 0.031 |
|  | ±0.008 | ±0.004 | ±0.007 | ±0.005 | ±0.006 | ±0.005 | ±0.006 | ±0.007 | ±0.008 |

**Table 3.** Calibrated indices for the observed Galactic globular clusters. The values are given in magnitude units.

## 4. Metallicity Calibrations

The $Mg_2$ index has been widely used as a metallicity indicator for composite stellar systems (Faber 1973, Burnstein 1979, BFGK84, Worthey et al. 1992). Actually, the definite relation between the $Mg_2$ index and the metal content $[Fe/H]$ is still under debate. Among the most recent results there are the calibrations of $Mg_2$ and other indices by BH90, the evolutionary synthesis models by Buzzoni et al. (1992), and the calibration by Barbuy (1994).

Many metallicity sensitive parameters are reported in the literature, derived from large band color indices (see the discussions in Covino et al. 1994), narrow-band photometry (Zinn & West 1984), and features in the color-magnitude diagrams of GCs (Peterson 1986). All these parameters were calibrated by high-resolution spectroscopy.

Our data and those of BFGK84 for 17 GGCs were used in Fig. 2 to show the relations $Mg_2$ versus $(U - B)_0$, $Q_{39}$, $(B - V)_{0,g}$, and $\Delta V_{1.4}$. Table 4 reports formal best fits between $Mg_2$ and these metallicity parameters. The relations are rather good and the error on the estimate is always of the order of the expected observational error. Our data and those of BFGK84 do not show any systematic difference both in slopes and intercepts.

Once metallicity effects in a stellar population can be singled out through a combined study of spectral features, it is possible to get a rather accurate analysis of the primeval nucleosynthesis products and the subsequent metal enrichment in stars of composite stellar systems.

The metallicity scale for GGCs more widely used is that by Zinn & West (1984). It is primarily based on the measure of some narrow-band photometric indices ($Q_{39}$) sensitive to the ultraviolet blanketing in the integrated light of GCs. The absolute scale is still uncertain at the level of a few tenths dex in $[Fe/H]$ but the relative ranking should be fairly accurate. Therefore, we have derived accurate relations between $[Fe/H]$ on Zinn & West's (1984) metallicity scale and homogeneous indices available for 26 GGCs of our and BFGK84 sample. For the clusters in common, we have taken a mean value weighted on the observational error usually quite comparable for our and BFGK84 sample. Each index shows a high degree of correlation (Fig. 3).

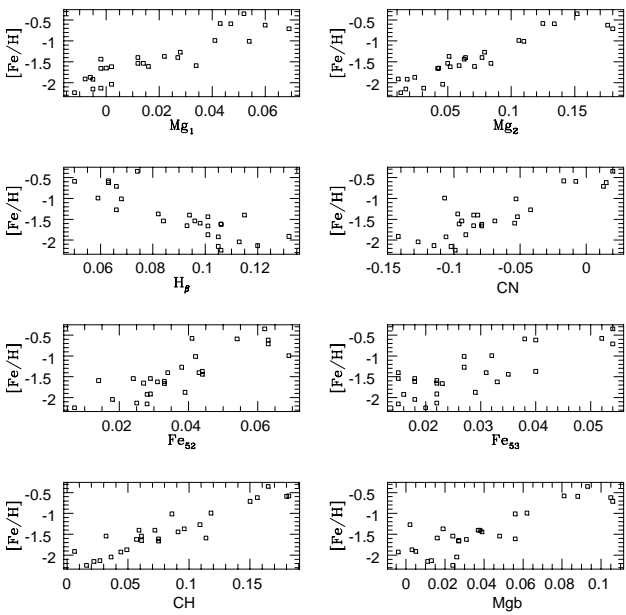

**Fig. 3.** Metallicity indices vs. $[Fe/H]$ on Zinn & West's (1984) scale. Each index shows a high degree of correlation.

|  | $m$ | $q$ | $\sigma$ |
|---|---|---|---|
| $(U-B)_0 - Mg_2$ | 2.311 | -0.060 | 0.050 |
| $Q_{39} - Mg_2$ | 2.608 | -0.057 | 0.055 |
| $(B-V)_{0,g} - Mg_2$ | 1.900 | 0.671 | 0.050 |
| $\Delta V_{1.4} - Mg_2$ | -7.082 | 3.023 | 0.050 |

**Table 4.** Least square fits of $Mg_2$ index vs. other metallicity parameters.

BH90 also determined the relations between their indices and the metal content $[Fe/H]$. These results are a fundamental step to estimate the metallicity of far stellar populations for which we can, usually, have information only from their integrated stellar content. Following the same procedure as in BH90, we determined least-square fits for our sample of 26 clusters. The obtained calibrations are given in Table 5. The error on the estimate is always very low, showing that Zinn & West's metallicity scale is completely consistent with our and BFGK84 measures.

| Index | $m$ | $q$ | $\sigma$ |
|---|---|---|---|
| $CH$ | 9.55 | -2.22 | 0.04 |
| $Mgb$ | 14.03 | -1.99 | 0.05 |
| $Fe_{52}$ | 28.20 | -2.47 | 0.06 |
| $Fe_{53}$ | 35.61 | -2.44 | 0.06 |
| $NaD$ | 9.05 | -2.07 | 0.08 |
| $H_\beta$ | -20.94 | +0.48 | 0.06 |
| $CN$ | 10.30 | -0.70 | 0.05 |
| $Mg_1$ | 19.88 | -1.81 | 0.04 |
| $Mg_2$ | 10.17 | -2.17 | 0.04 |

**Table 5.** Calibrations of metallicity indices vs. metal content $[Fe/H]$.

## 5. Synthetic Indices

The study of metallicities and abundance ratios in composite stellar systems can trace their chemical evolution and give information on their stellar populations and ages.

Buzzoni et al. (1992, 1994) have studied the dependence of some indices ($Mg_2$, $Fe_{52}$, $Fe_{53}$, and $H_\beta$) on the primary stellar parameters in order to establish, when possible, analytical formulas describing the dependence of the indices on $T_{eff}$, $\log g$, and $[Fe/H]$ for single stars. Since the calibration for GCs is the first step to perform, in order to derive the relation between metallicity parameters and $[Fe/H]$, special attention was devoted to Simple Stellar Populations (SSPs), which are aggregates of coeval single stars with an homogeneous metal content. As well known, and shown also recently by Bressan et al. (1993), SSPs can be considered the "building blocks" of composite stellar systems.

Following Buzzoni et al. (1992), if we know values of some index for a large enough number of single stars, the synthetic index for the whole aggregate can be easily computed. Calculations were performed by Buzzoni for a large number of models of SSPs using his code for evolutionary population synthesis (Buzzoni 1989). Global index values were determined for aggregates of different ages, Initial Mass Function (IMF) slopes, and mass loss parameters following the prescription by Reimers (1975).

In Fig(s). 4 and 5 we compare our and BFGK84 mean weighted indices, as in Fig. 3, with models computed for 15 Gyr old SSPs. The Salpeter value, 2.25, has been assumed for the IMF slope and 0.5 for the mass loss parameter (see Buzzoni 1989 and Covino et al. 1994). The HB morphologies adopted in this paper were attributed to the clusters with the same criteria used in Covino et al. (1994).

On the plane $Mg_2 - Fe_{52}$ there is not a sensitive difference between R- and I-HB morphology models. Moreover,

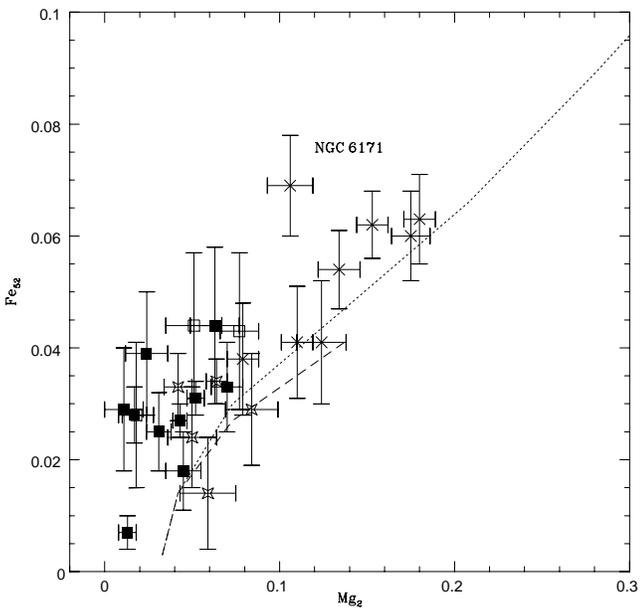

**Fig. 4.** SSP models by Buzzoni (1989) ($\tau = 15$ Gyr, $s = 2.35$, and $\eta = 0.5$) compared with observational data on the $Mg_2 - Fe_{52}$ plane. Crosses, stars, and solid squares refer to R-, I-, and B-HB morphology respectively. An open square is used for clusters for which the HB morphology is unknown. Dotted and dashed lines refer to R-HB and I-HB models respectively.

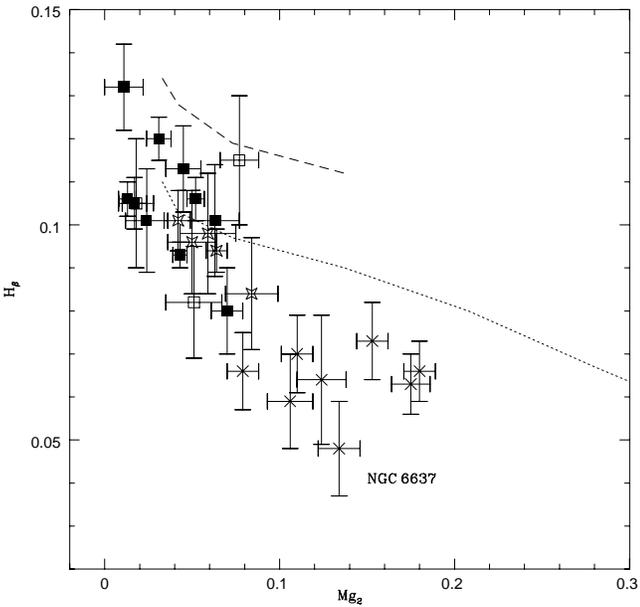

**Fig. 5.** Same as figure 4 but on the $Mg_2 - H_\beta$ plane.

as these indices concern, from an intermediate morphology (Buzzoni et al. 1994). Actually, the change of the color distribution on the HB should affect more the $H_\beta$ index than the iron or magnesium indices, as discussed later on.

Even if the global trend is rather well traced by the models, the observational points appear to lie above the theoretical lines. A formal least-squares fit between $Mg_2$ and $Fe_{52}$ for observed data and for theoretical predictions in the linear part (i.e. $Mg_2 \geq 0.05$) shows that the fiducial line traced by our and BFGK84 data has a slightly lower slope and a positive shift with respect to the models. The relations are as follows: $Fe_{52} = 0.253 \cdot Mg_2 + 0.019$ for observed data and $Fe_{52} = 0.306 \cdot Mg_2 + 0.005$ for theoretical predictions. Actually, the maximum scatter between models and observational indices reaches about $0.03 \div 0.04$ mag in $Fe_{52}$ or $\sim 0.1$ mag in $Mg_2$ for the cluster NGC 6171. Observed clusters with lower metal content, i.e. B- and I-HB clusters, show a less definite trend and lie systematically over the model fiducial line. This can be attributed to an underestimation of $Fe_{52}$ index by the models for the lowest metal contents, and/or to the well known difficulty in determining pseudo-equivalent widths for weak features in integrated spectra.

| Index | m | q |
|---|---|---|
| $Fe_{52}$ | 26.38 | -2.37 |
| $H_\beta$ | -23.21 | +1.51 |
| $Mg_2$ | 7.25 | -2.02 |

**Table 6.** Calibrations obtained from theoretical SSPs of Buzzoni et al. (1992, 1994). The relation for $H_\beta$ was computed considering data from I-HB models for lower and intermediate metallicity and for R-HB for the highest metallicity. That for $Mg_2$ was determined for $[Fe/H] \geq -2.00$ since at the lowest metal contents the relation is not linear.

$H_\beta$ fiducial lines, as pointed out by Buzzoni et al. (1994), show clearly the effect of different color distribution on the HB at least for the I- and R-HB morphologies. On the contrary, B-HB models are quite similar to I-HB models. The observed data do not show a good agreement with models and the observational points are systematically lower than the theoretical predictions. We can also compare the relations between metal content and indices reported in Table(s) 5 and 6 for observational and theoretical data respectively. The fit between $[Fe/H]$ and $H_\beta$ was determined using models with I-HB up to a metal content of $[Fe/H] = -1$ and models with R-HB for higher metallicity.

Since $H_\beta$ is basically an indicator of the temperature of the upper MS, a way to decrease the values computed by models is to move the upper MS to lower temperatures. This can be performed by increasing substantially the age relations $[O/Fe] = -0.5[Fe/H]$ for $[Fe/H] > -1$, and $[O/Fe] = -0.2[Fe/H] + 0.3$ for lower metallicities, reported by Buzzoni et al. (1994), an enhancement of $[O/Fe]$ can decrease $H_\beta$ values by roughly 0.3 Å or just 0.02 mag at $[Fe/H] \simeq -1$. These relations can slightly overestimate the actual amount of $[O/Fe]$, as suggested by Minniti et al. 1993. Taking into account this improvement, the maximum scatter between observational points and theoretical predictions, for all the HB morphologies, results about $0.02 \div 0.03$ magn.

In any case, R-HB clusters show observed $H_\beta$ indices around a mean value of about $\simeq 0.06$, suggesting that this index is not very sensitive to the metal content in the range of the highest metallicities.

## 6. Statistical Dimensionality

Almost all spectrophotometric indices are defined as metallicity indicators. In the previous sections, we have seen that each index considered in this work can provide a reliable measure of the global metal content of a composite stellar system. A different approach to investigate on the reliability of the relations between spectrophotometric indices and metal content is given by the use of a sophisticated statistical analysis as the Principal Components Analysis (PCA). An introduction on multivariate statistical methods can be found in Kurtz (1983) and Murtagh & Heck (1987), while more detailed mathematical descriptions are provided by Mardia et al. (1989).

which converts any set of variables into a new one, the Principal Components (PCs), which are uncorrelated. The linear transformation of the original variables allows one to simplify the structure of the correlation (or covariance) matrix in order to make the interpretation of the data more straightforward. PCA can also help to visualize the data by the selection of the most important planar views.

The set of our data includes 9 indices: $CH$, $Mgb$, $Fe_{52}$, $Fe_{53}$, $NaD$, $H_\beta$, $CN$, $Mg_1$, and $Mg_2$, obtained from our and BFGK84 samples. In total we have considered 26 GGCs (Table 3 and BFGK84 Table 2A).

| Eigenvalues | As Percentages | Cumul. Percentages |
|---|---|---|
| 6.7927 | 75.4741 | 75.4741 |
| 0.7451 | 8.2794 | 83.7535 |
| 0.4297 | 4.7745 | 88.5281 |
| 0.3713 | 4.1253 | 92.6534 |
| 0.3139 | 3.4877 | 96.1411 |
| 0.1385 | 1.5391 | 97.6801 |
| 0.1167 | 1.2970 | 98.9771 |
| 0.0811 | 0.9015 | 99.8786 |
| 0.0109 | 0.1214 | 100.0000 |

**Table 7.** Percentage of variance explained by the principal components.

The PCA on the correlation matrix has provided the eigenvalues reported in Table 7. Of course, just one component is able to describe almost all the variance contained in the data as the second and even more the third components are only marginally meaningful.

| Index | $PC_1$ | $PC_2$ | $PC_3$ |
|---|---|---|---|
| $CH$ | -0.36 | +0.16 | -0.03 |
| $Mg_b$ | -0.34 | +0.24 | +0.16 |
| $Fe_{52}$ | -0.33 | -0.26 | +0.13 |
| $Fe_{53}$ | -0.32 | -0.18 | +0.69 |
| $NaD$ | -0.24 | -0.86 | -0.19 |
| $H_\beta$ | +0.33 | -0.14 | +0.59 |
| $CN$ | -0.34 | +0.17 | +0.11 |
| $Mg_1$ | -0.36 | +0.04 | -0.27 |
| $Mg_2$ | -0.37 | +0.19 | -0.03 |

**Table 8.** Weights on the first three components for each index involved in the analysis.

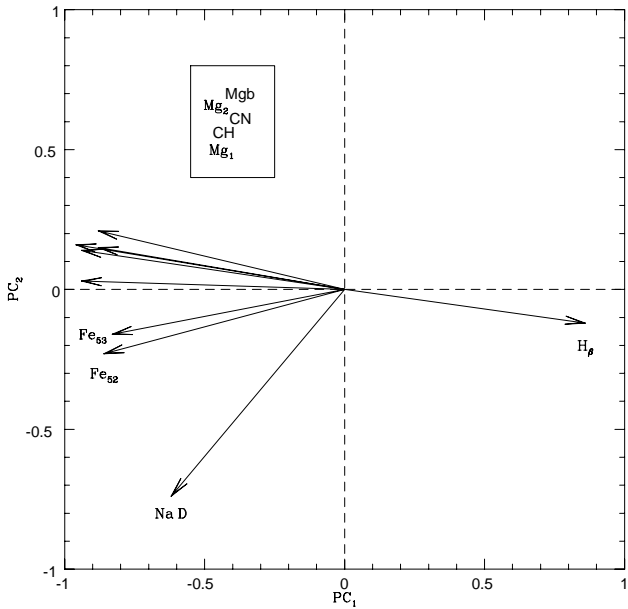

**Fig. 6.** Correlation of indices between the first two principal components.

The eigenvectors associated to the first three PCs are reported in Table 8. The most interesting result is that the first PC is substantially a simple mean of each index considered. The weights are all comparable with a slight portant index, with respect the first PC, is the sodium index $NaD$. This can be due to the considerable effect of the interstellar sodium line as already shown in the fourth panel of Fig. 1 and previously reported by Cohen (1973, 1975). In fact, almost all the second component is due to the $NaD$ index.

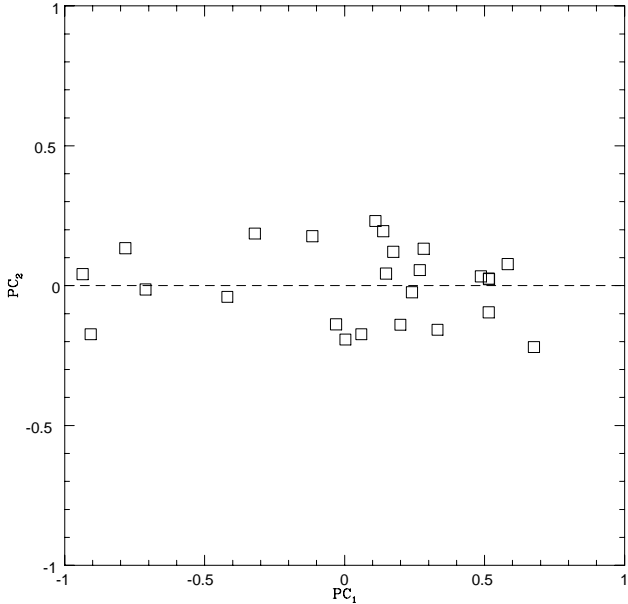

**Fig. 7.** Plane determined by the first two principal components.

The same conclusions can be achieved considering Fig. 6 where the projections of the vectors on the axes visualize the correlation of each index with the first and second PCs. Only the sodium index needs a second parameter to be described over the observational errors. The diagram shows also that $H_\beta$ index is anticorrelated to the others, as it is well known.

Of course, we can identify the first parameter (PC) with the metallicity as inferred also by Fig. 3. If we perform the PCA on the same sample but removing the sodium index, we would obtain a full mono-parametric family of indices as shown in Fig. 7.

### 7. Conclusions

Spectrophotometric indices for 18 Galactic Globular clusters, obtained from CCD observations and careful reductions, were used to determine reliable calibrations on metallicity $[Fe/H]$ in the full range covered by GGCs. We chose to measure the indices in the bandpasses adopted by Burnstein et al. (1984) in order to calibrate our data in an homogeneous and reproducible system. Our observations to obtain an homogenous sample of spectrophotometric indices for 26 GGCs.

Relations with indices defined in different bandpasses or resolutions and with metallicity photometric indices or parameters were computed. Among the various metallicity parameters, we considered $(U-B)_0$, $Q_{39}$, $(B-V)_{0,g}$, and $\Delta V_{1.4}$. In each case we obtained satisfactory correlations.

New calibrations for a large number of indices versus $[Fe/H]$ were also computed. They cover the whole range of metallicity observed in GGCs and are determined for clusters with various physical parameters. The relations are consistent with Zinn & West's (1984) metallicity scale.

Observational data were compared with synthetic indices derived from Buzzoni's (1989) models to test the theoretical predictions and their applications to the studies of external galaxies. Detailed discussions have been also performed for $Mg_2$, $Fe_{52}$, and $H_\beta$ observed indices.

Clusters with B- and I-HB morphology do not show a well defined sequence on the plane $Mg_2/Fe_{52}$, while the $H_\beta$ index seems to loose sensitivity at the highest metal contents assuming a roughly constant value of $\sim$ 0.06. Even if the global trends are rather well traced by the models, they do not fit the observational points. Moreover, the scenario confirms Buzzoni's et al. (1994) deduction that a certain degree of oxygen enhancement is necessary to have a better agreement between observations and theoretical predictions. With the enhancement suggested by these Authors, the lowering on $H_\beta$ index results about 0.02 mag.

Finally, some results were also discussed from a statistical point of view by the aid of a principal component analysis. We quantified the sensitivity of spectrophotometric indices as metallicity indicators for GGCs. GGC indices define a full mono-parametric family just depending on the metal content.

*Acknowledgments* The Authors are indebted to prof. L. Mantegazza to have made available his observational unpublished material. SC and SG want to thank the Osservatorio Astronomico di Brera-Milano for the kind hospitality.